\documentclass[preprint,showpacs,preprintnumbers,amsmath,amssymb,endfloats]{revtex4}
\usepackage{amsmath,graphicx,dcolumn}

\newcommand{\AFeAs}{AFe$_2$As$_2$}
\newcommand{\BaNiAs}{BaNi$_2$As$_2$}
\newcommand{\SrNiAs}{SrNi$_2$As$_2$}
\newcommand{\EuNiAs}{EuNi$_2$As$_2$}
\newcommand{\ANiAs}{ANi$_2$As$_2$}

\newcommand{\ThCrSi}{ThCr$_2$Si$_2$}

\begin{document}

\title{Superconductivity in SrNi$_2$As$_2$ Single Crystals}

\author{E. D. Bauer}
\affiliation{Los Alamos National Laboratory, Los Alamos, New Mexico 87545, USA}
\author{F. Ronning}
\affiliation{Los Alamos National Laboratory, Los Alamos, New Mexico 87545, USA}
\author{B. L. Scott}
\affiliation{Los Alamos National Laboratory, Los Alamos, New Mexico 87545, USA}
\author{J. D. Thompson}
\affiliation{Los Alamos National Laboratory, Los Alamos, New Mexico 87545, USA}

\date{\today}

\begin{abstract}
The electrical resistivity $\rho(T)$ and heat capacity $C(T)$ on single crystals of
\SrNiAs{} and \EuNiAs{} are reported. While there is no evidence for a structural
transition in either compound, \SrNiAs{} is found to be a bulk
superconductor at $T_c=0.62$ K with a Sommerfeld coefficient
of $\gamma$= 8.7 mJ/mol K$^2$ and a small upper critical field $H_{c2} \sim 200$ Oe. No superconductivity was found in
\EuNiAs{} above 0.4 K, but anomalies in $\rho$ and $C$ reveal that magnetic order
associated with the Eu$^{2+}$ magnetic moments occurs at $T_m $=14 K.
\end{abstract}


\maketitle

The \ThCrSi{} structure type is well known for accommodating a
superconducting ground state, particularly in the heavy fermion
community with superconductors such as CeCu$_2$Si$_2$
 and URu$_2$Si$_2$.\cite{Hess93} Soon after
the discovery of superconductivity in LaFeAsO$_{1-x}$F$_x$ at $T_c$
= 26 K with the structure type ZrCuSiAs,\cite{KamiharaJACS2008} it
was realized that the related compounds (\AFeAs{} with A = Ba, Sr,
Ca, Eu) in the \ThCrSi{} structure are also superconducting either
with doping\cite{Rotter2008b,GFChen2008b,Sasmal2008,Wu2008Ca,Jeevan2008EuFe2As2SC}
or under pressure.
\cite{Park2008CaFe2As2,Torikachvili2008CaFe2As2,Alireza2008BaFe2As2}.
While systems with Fe$_2$As$_2$ planes have the highest $T_c$'s
to date, superconductivity has been found in both structure types
 with either Ni$_2$P$_2$
(Refs. \onlinecite{Watanabe2007LaNiPO,Mine2008BaNi2P2}) or Ni$_2$As$_2$
(Refs. \onlinecite{Watanabe2008LaNiAsO,
Fang2008NiAs,Li2008NiAs,RonningJPCM2008BaNi2As2}) layers.

Here we report the observation of superconductivity in single
crystals of \SrNiAs{} at $T_c$ = 0.62 K, as determined by heat
capacity, in the absence of a structural phase transition (below 400 K). Following our initial observation of superconductivity in \BaNiAs,\cite{RonningJPCM2008BaNi2As2} this represents the second superconducting system in
the \ThCrSi{} structure with Ni$_2$As$_2$ layers. In addition, we report that our EuNi$_2$As$_2$
single crystals grown from Pb flux are not superconducting above
0.4 K.

Single crystals of SrNi$_2$As$_2$ and EuNi$_2$As$_2$
were grown in Pb flux in the ratio (Sr,Eu):Ni:As:Pb=1:2:2:20.
The starting elements were placed in an alumina crucible and sealed
under vacuum in a quartz ampoule. The ampoule was placed in a
furnace and slowly heated 1050 $^{\circ}$C, as described in Ref. \onlinecite{RonningJPCM2008BaNi2As2}.  The sample was then cooled slowly ($5^{\circ}$C hr$^{-1}$) to 600 $^{\circ}$C, at which point the excess
Pb flux was removed with the aid of a centrifuge. For \SrNiAs{} the
resulting plate-like crystals were heavily imbedded in a yet unidentified needle-like
impurity phase. From single crystal x-ray refinements,  the plate-like samples were confirmed to crystallize in the
ThCr$_2$Si$_2$ tetragonal structure (space group no. 139,
I4/mmm). The refinement for \SrNiAs{} [R(I$>$2$\sigma$) = 3.7\%] at
124 K yields lattice parameters $a$ = 4.1374(8) \AA{} and $c$ =
10.188(4) \AA{} and fully occupied ($>$98\%) atomic positions Sr
2a(0,0,0), Ni 4d(0.5,0,0.25) and As 4e(0,0,z) with z = 0.3634(1)
consistent with previous
reports.\cite{Mewis1980,Pfisterer1980,Pfisterer1983} The refinement
for \EuNiAs{} [R(I$>$2$\sigma$) = 5.09\%] at 124 K gives lattice
parameters $a$ = 4.0964(6) \AA{}, $c$ =  10.029(3) \AA{} and fully
occupied ($>$98\%) atomic positions Eu 2a(0,0,0), Ni 4d(0.5,0,0.25)
and As 4e(0,0,z) with z = 0.3674(2) also consistent with previous
reports.\cite{JeitschkoJLCM1988MNi2As2,GhadraouiMRB1988LnNi2As2}

\begin{figure}
\includegraphics[width=3.3in]{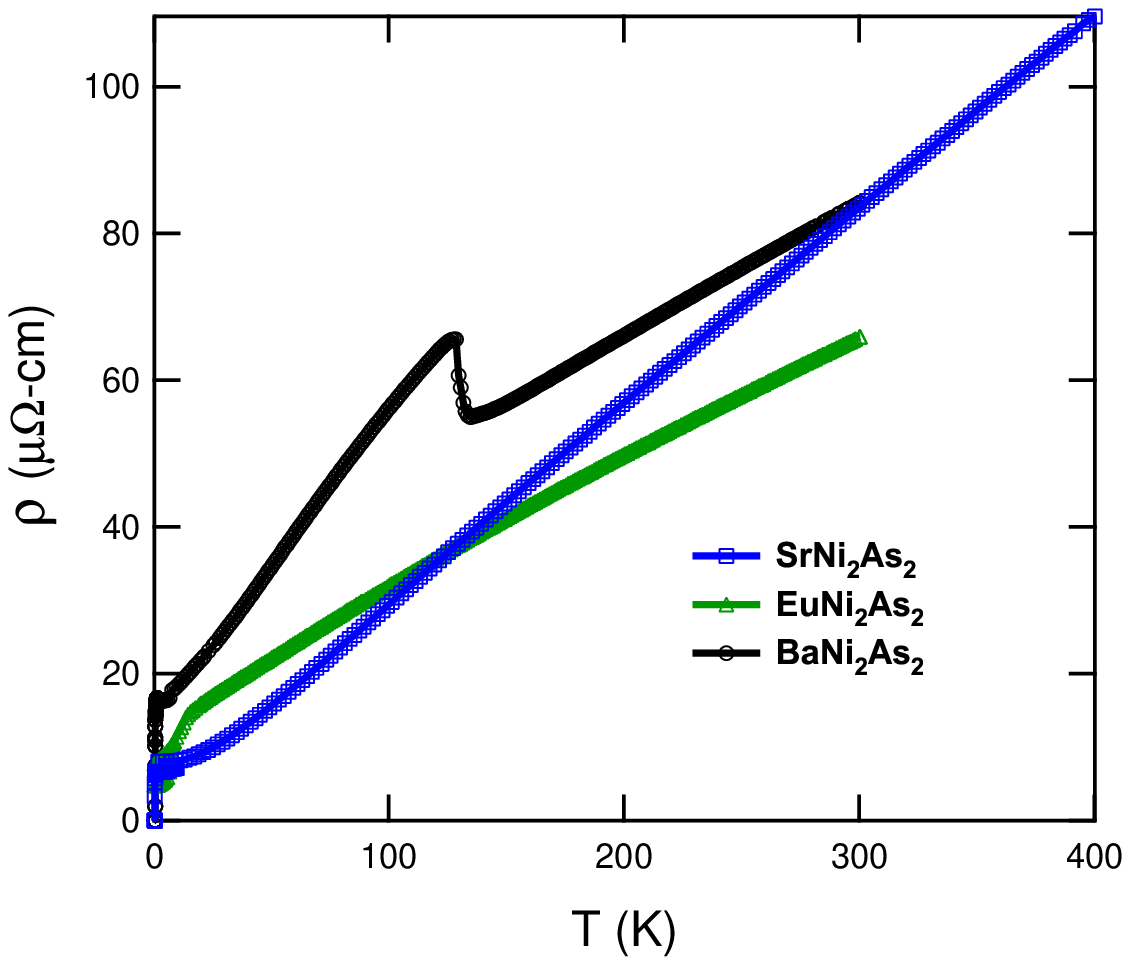}
\caption{(color online) In-plane electrical resistivity
$\rho(T)$ (I$\parallel$ab) for selected ANi$_2$As$_2$ (A=Ba, Sr, Eu) compounds.}
\label{SrNiAsRes}
\end{figure}

\begin{figure}[htbp]
     \centering
     \includegraphics[width=0.5\textwidth]{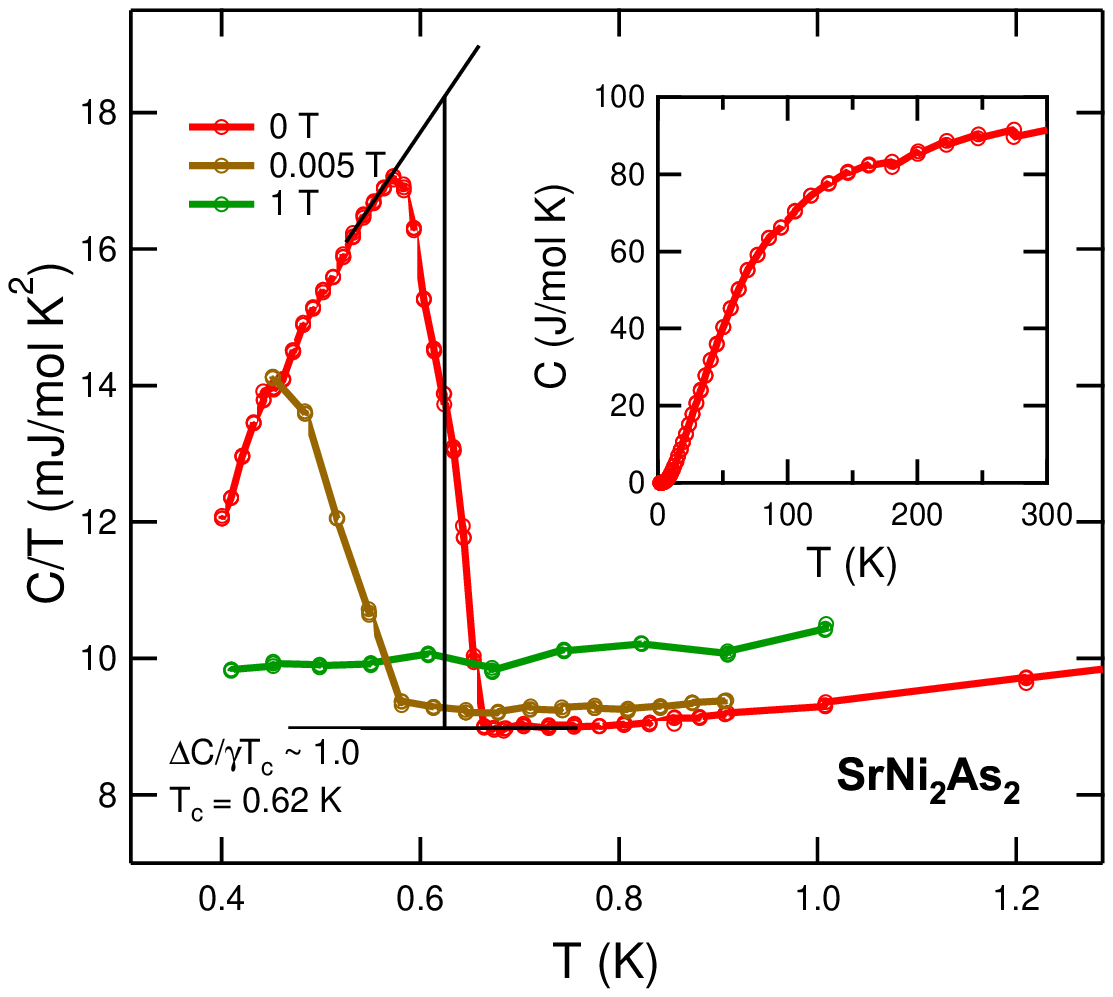}
     \caption{(color online) Low temperature specific heat C
versus temperature T of SrNi$_2$As$_2$ \cite{Cpfootnote} for various magnetic fields ($H||c$).
The inset displays
the high temperature heat capacity with no indication of a first
order phase transition at higher temperatures.}
\label{Cp}
 \end{figure}

The in-plane electrical resistivity data for \ANiAs{} (A=Ba, Sr, Eu) is
shown in Fig. \ref{SrNiAsRes}. All samples exhibit metallic behavior. \SrNiAs{} is a relatively good metal with a RRR (= $\rho$(300 K)/$\rho$(4 K)) of 11, and a residual resistivity of 7
$\mu\Omega$-cm. The resistivity of \EuNiAs{}
 exhibits a kink at $T_m=14 $ K associated with magnetic ordering of the Eu$^{2+}$ moments,
consistent with previous reports.\cite{GhadraouiMRB1988LnNi2As2}
For \SrNiAs, there is no evidence of a structural transition below 400 K , in contrast to \BaNiAs, which has a clear first order transition at $T_0= 130$ K.
The lack of a phase transition in \SrNiAs{} is also
provided by the heat capacity data shown in the inset of Fig.
\ref{Cp}.\cite{Cpfootnote}

\begin{figure}
\includegraphics[width=3.3in]{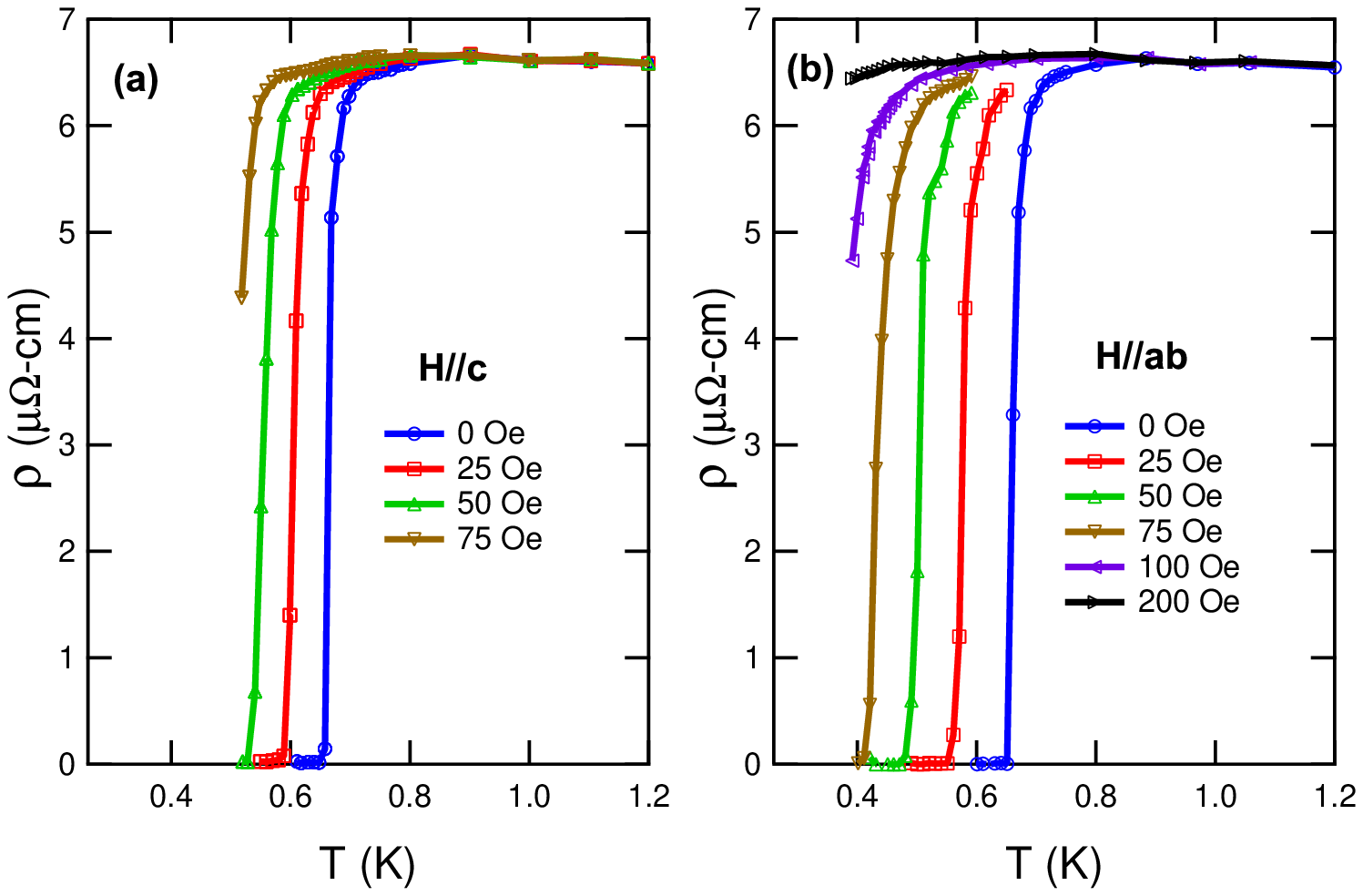}
\caption{(color online) Electrical  resistivity $\rho(T)$ of \SrNiAs{} showing the superconducting transition for $H \parallel \emph{c}$ (a) and
     $H \parallel \emph{ab}$ (b). The current was maintained perpendicular to the
     magnetic field.}
\label{SCRes}
\end{figure}

Figure \ref{SCRes} presents the low temperature in-plane resistivity
data for \SrNiAs{} with fields applied parallel and perpendicular to
the c-axis. In zero field, a sharp superconducting transition is
observed at $T_c$ = 0.66 K, defined as the mid-point resistive anomaly. With increasing magnetic field, the transition remains sharp and is quickly
suppressed. The specific heat shown in Fig.  \ref{Cp} confirms the bulk
nature of superconductivity in \SrNiAs. The zero resistance state
coincides exactly with the onset of the specific heat transition,
from which we extract a superconducting transition temperature of
$T_c= 0.62$ K by an equal area construction. A fit to
the data from 0.7 K to 3 K of $C/T = \gamma + \beta{}T^2 +
\delta{}T^4$,  a Sommerfeld coefficient $\gamma$ = 8.7
mJ/mol-K$^2$ is obtained. Using this value, the
ratio of the specific heat jump at $T_c$ to the electronic specific heat is estimated to be $\Delta{}C/\gamma{}T_c \simeq$ 1.0.   From the $\beta$
coefficient = 0.67 mJ/molK$^4$, one obtains a Debye temperature
$\Theta_D$ = 244 K.

\begin{figure}[htbp]
     \centering
     \includegraphics[width=0.5\textwidth]{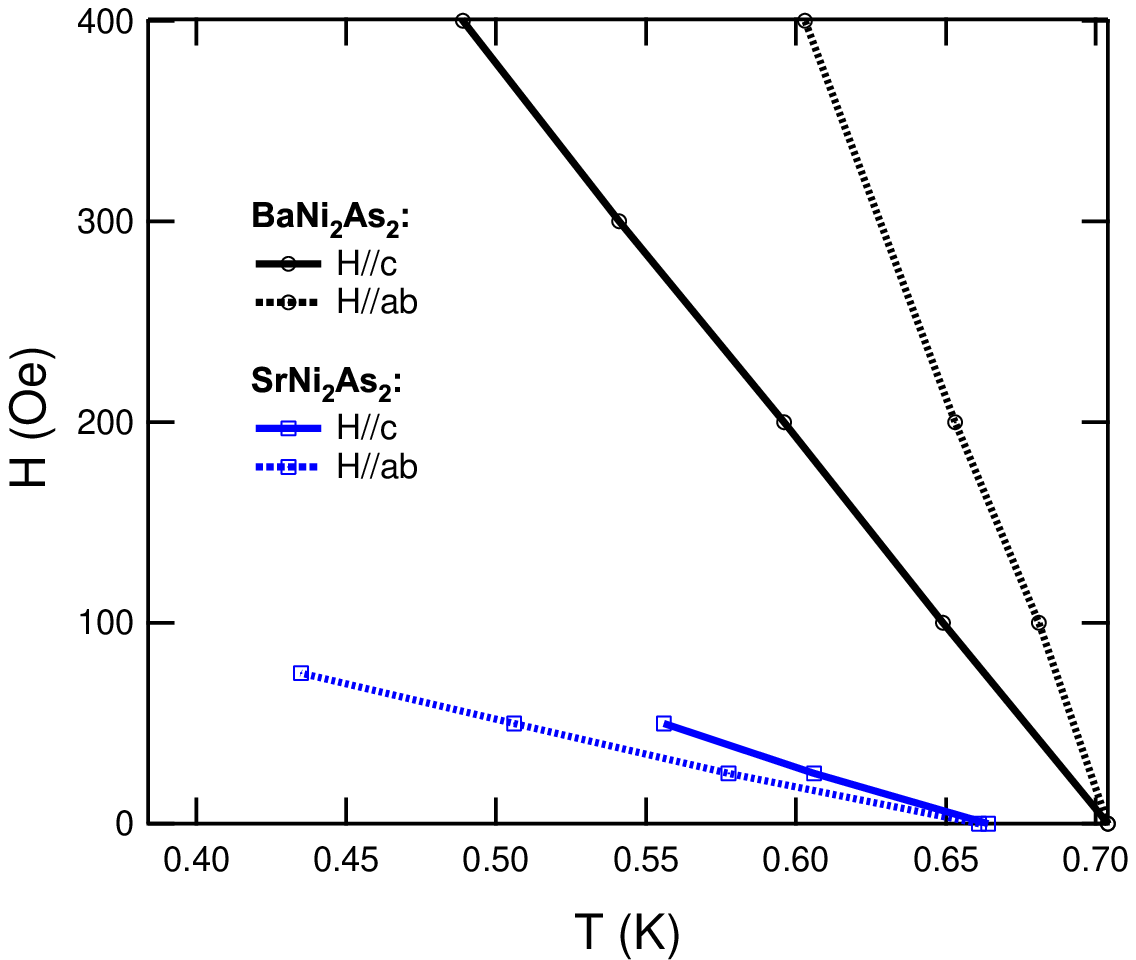}
     \caption{(color online)  Magnetic field-temperature $H-T$ phase diagram of \SrNiAs. The  upper critical field  $H_{c2}$ for
     $H \parallel \hat{c}$ and
     $H \parallel \hat{ab}$ was determined by the resistive midpoint in Fig. \ref{SCRes}(a)
     and (b). The data for \BaNiAs{} from Ref. \onlinecite{RonningJPCM2008BaNi2As2} is for the
     zero resistance state.} \label{Hc2vsT}
 \end{figure}

The magnetic field-temperature $H-T$ phase
diagram of \SrNiAs{} is shown in Fig. \ref{Hc2vsT} determined from the $\rho(T)$ curves in Fig. \ref{SCRes}, along with the data for
\BaNiAs{}  for
comparison.\cite{RonningJPCM2008BaNi2As2} For \SrNiAs,  the zero-temperature orbital
critical field\cite{WHH1966} $H_{c2}^*$(0) = -0.7$\,T_c\,dH_{c2}/dT_c$  is determined to
be 210 Oe and 150 Oe for $H\parallel{}c$ and $H\parallel{}ab$,
respectively. From this, the superconducting coherence length is estimated via $H_{c2}^*(0) = \Phi_0/2 \pi \xi_0^2$,\cite{Tinkham75} yielding
$\xi_0^{ab} = 1477$ \AA, and $\xi_0^{c}= 1250$ \AA. Values for the Fermi velocity $v_F^{ab}=
7.1 \times 10^6$ cm/s and $v_F^{c}= 6.0 \times 10^6$ cm/s  are obtained from $\xi_0= 0.18 \hbar\, v_F /k_B\, T_c$.  Surprisingly, while the absolute value of $T_c$ in
zero magnetic field is very similar for the two compounds, the anisotropy ($H_{c2}^{ab}/H_{c2}^c$) of the upper critical field,
which is a factor of 2.1 in \BaNiAs{}, reverses sign ($H_{c2}^{c}/H_{c2}^{ab}$=1.4) in
\SrNiAs{} and the
overall magnitude of the upper critical field is nearly an order of magnitude
smaller.
The differences in the $H_{c2}$ between the two compounds may be due changes in electronic structure resulting from the structural transition in
\BaNiAs{} that is not present in \SrNiAs.

\begin{figure}[htbp]
     \centering
     \includegraphics[width=0.5\textwidth]{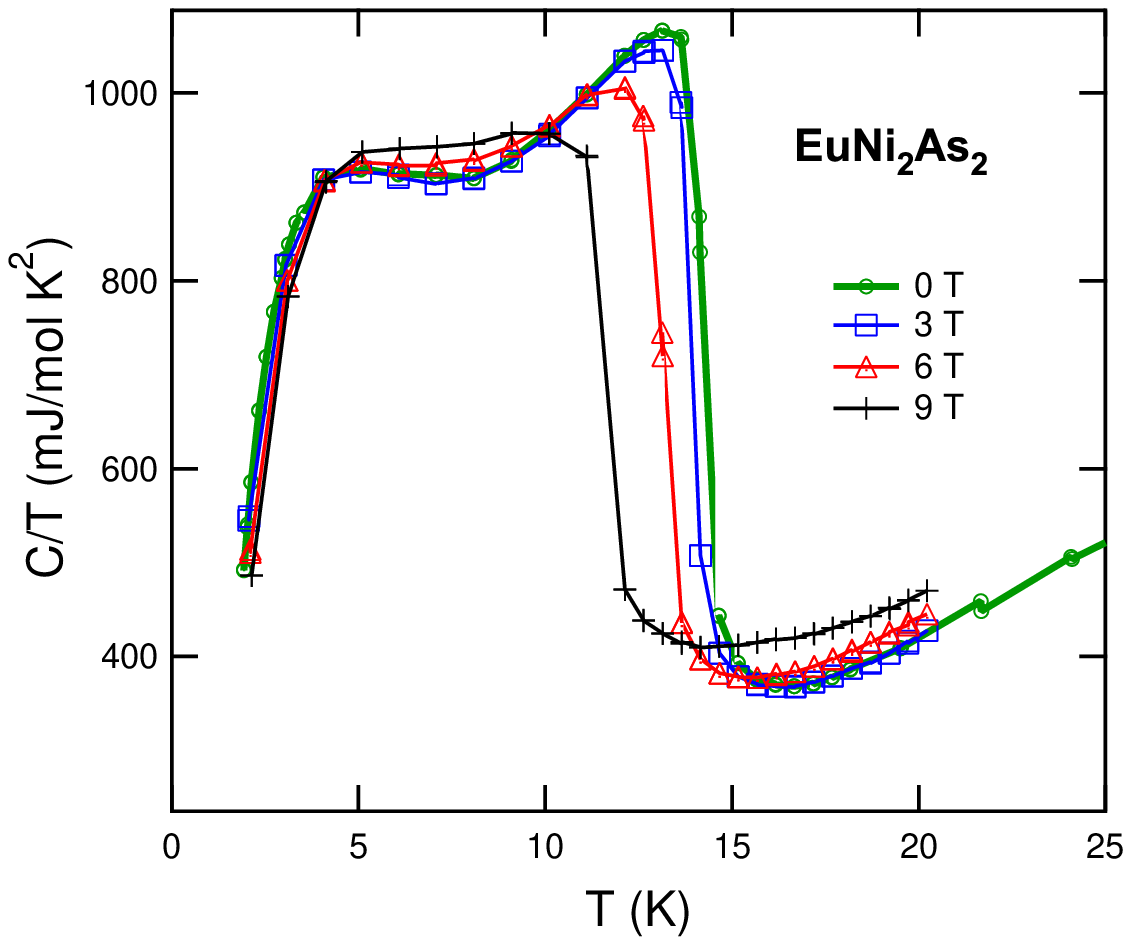}
     \caption{(color online) Heat capacity data
versus temperature for EuNi$_2$As$_2$ in zero and applied magnetic
field \cite{Cpfootnote}. The magnetic field was applied along the
c-axis.} \label{EuCp}
 \end{figure}

The specific heat, plotted as $C/T$, for \EuNiAs{} is  shown in Fig.  \ref{EuCp} in magnetic fields up to 9 T ($H||c$).
A sharp anomaly occurs at the magnetic ordering temperature $T_m=14$ K (consistent with the kink in $\rho(T)$, Fig. \ref{SrNiAsRes}), as well
as a broader hump at  $\sim$ 4 K.  A
magnetic field along the c-axis modestly suppresses the transition,  consistent with previous reports of antiferromagnetic ordering in polycrystalline samples.\cite{GhadraouiMRB1988LnNi2As2,Raffius1993}   There is no indication of superconductivity
above 0.4 K in \EuNiAs.

It is interesting that superconductivity with very similar
transition temperatures is found both in
\BaNiAs\cite{RonningJPCM2008BaNi2As2} and in \SrNiAs, despite the
differences in structural parameters caused by the smaller Sr$^{2+}$
ions, as well as the presence of a first order structural transition in
\BaNiAs{} that is  possibly also magnetic.  Recent theoretical work by Subedi and coworkers\cite{Subedi08} indicate that the superconducting properties of the related Ni-analog LaNiPO may be explained within a conventional electron-phonon approach, yielding a low value of $T_c=2.6$, consistent with experiment; the authors go on to suggest that the Fe-As superconductors may be in a separate class from their Ni-based counterparts.  However, a scenario  has been put forth by Cvetkivoc and Tesanovic\cite{Cvetkivoc08}  involving a multiband Fermi surface in the layered FeAs superconductors to produce the large values of $T_c$ may also be an appropriate description of these ANi$_2$As$_2$ superconductors as well. Further work is in progress to elucidate the nature of the superconductivity in these Ni-based materials and its relation to fine details of the electronic structure.

In conclusion, specific heat and electrical resistivity measurements \SrNiAs{} single crystals  reveal  bulk superconductivity at 0.62 K, which shows no sign of a structural/magnetic anomaly below 400 K. Magnetic ordering associated with the Eu magnetic moments is observed in single crystalline \EuNiAs.  No evidence
for superconductivity is observed in this compound above 0.4 K.

\begin{acknowledgments}
Work at Los Alamos National Laboratory was
performed under the auspices of the U.S. Department of Energy.
\end{acknowledgments}


\begin{thebibliography}{25}
\bibitem[{\citenamefont{}()}]{Cpfootnote}
  \bibinfo{journal}{Due to the small size of the heat capacity signal, we
  are unable to properly subtract the addenda. This is particularly
  evident at high temperatures, where we do not recover the Dulong-Petit value.
  We believe this is also responsible for the shift of approximately 1 mJ/mol K$^2$
  at low temperatures and a magnetic field of 1 T.}
\end{thebibliography}

\end{document}